%% file: main.tex
\documentclass[10pt,conference]{IEEEtran}
\IEEEoverridecommandlockouts
\usepackage{cite}
\usepackage{amsmath,amssymb,amsfonts}
\usepackage{algorithmic}
\usepackage{graphicx}
\usepackage{float}
\usepackage{comment}
\usepackage{textcomp}
\usepackage{xcolor}
\usepackage{xspace}
\usepackage{balance}
\usepackage{braket}
\usepackage[section]{placeins}
\usepackage{array}
\usepackage{makecell}
\usepackage{dblfloatfix}   
\usepackage{placeins}      
\usepackage[ruled,vlined,linesnumbered]{algorithm2e}

\usepackage[ruled,vlined]{algorithm2e}
\definecolor{darkgreen}{rgb}{0,0.6,0}
\def\BibTeX{{\rm B\kern-.05em{\sc i\kern-.025em b}\kern-.08em
    T\kern-.1667em\lower.7ex\hbox{E}\kern-.125emX}}

\newcommand{\eg}{\emph{e.g.,}\xspace}

\newcommand{\descStep}[2]{\noindent \textbf{#1: } #2}

\usepackage{soul} 

\usepackage{tikz}
\usepackage{graphicx}   
\usetikzlibrary{arrows.meta,positioning,calc}
\tikzset{
  gscnode/.style={
    rectangle, draw, rounded corners, align=center,
    font=\scriptsize, inner sep=2.5pt,
    text width=2.35cm, minimum height=0.72cm
  },
  gscarrow/.style={-{Latex[length=1.5mm]}, thick}
}

\begin{document}

\title{GSC-QEMit: A Telemetry-Driven Hierarchical Forecast-and-Bandit Framework for Adaptive Quantum Error Mitigation\\
}

\author{
\IEEEauthorblockN{
Steven Szachara\IEEEauthorrefmark{1},
Sheeraja Rajakrishnan\IEEEauthorrefmark{1},
Dylan Jay Van Allen\IEEEauthorrefmark{2},
Jason Pollack\IEEEauthorrefmark{2},
Travis Desell\IEEEauthorrefmark{1},
Daniel Krutz\IEEEauthorrefmark{1}
}
\IEEEauthorblockA{
\IEEEauthorrefmark{1}Department of Software Engineering, Rochester Institute of Technology, Rochester, USA\\
Emails: ss9270@rit.edu, sr8685@rit.edu, tjdvse@rit.edu, dxkvse@rit.edu
}
\IEEEauthorblockA{
\IEEEauthorrefmark{2}Institute for Quantum \& Information Sciences, Syracuse University, Syracuse, USA\\
Emails: djvanall@syr.edu, japollac@syr.edu
}
}

\maketitle

\input{2026/abstract}

\input{2026/intro}

\input{2026/relatedworks}
\vspace{-2pt}
\input{2026/methodology}

\input{2026/implementation}

\input{2026/results}

\input{conclusion}

\label{ch:references}

\balance
\bibliographystyle{IEEEtran}
\bibliography{references}

\end{document}

%% file: 2026/abstract.tex
\begin{abstract}
Quantum error mitigation (QEM) is essential for extracting reliable results from near-term quantum devices, yet practical deployments must balance mitigation strength against runtime overhead under time-varying noise. We introduce \emph{GSC-QEMit}, a telemetry-driven, \textbf{context--forecast--bandit} framework for \emph{adaptive} mitigation that switches between lightweight suppression and heavier intervention as drift evolves. GSC-QEMit composes three coupled modules: (G) a Growing Hierarchical Self-Organizing Map (GHSOM) that clusters streaming telemetry into operating contexts; (S) an uncertainty-aware subsampled Gaussian-process forecaster that predicts short-horizon fidelity degradation; and (C) a cost-aware contextual multi-armed bandit (CMAB) that selects mitigation actions via Thompson sampling with explicit intervention cost. We evaluate GSC-QEMit on benchmark circuit families (GHZ, Quantum Fourier Transform, and Grover search) under nonstationary noise regimes simulated in Qiskit Aer, using an instrumented testbed where action labels correspond to graded mitigation intensity. Across Clifford, non-Clifford, and structured workloads, GSC-QEMit improves average logical fidelity by \textbf{+9.0\%} relative to unmitigated execution while reducing unnecessary heavy interventions by reserving them for inferred noise spikes. The resulting policies exhibit a favorable fidelity--cost trade-off and transfer across the evaluated workloads without circuit-specific tuning.
\end{abstract}

%% file: 2026/intro.tex
\section{Introduction}
\label{sec:intro}
In the noisy intermediate-scale quantum (NISQ) regime, a central practical challenge is extracting reliable results from devices whose error processes drift over time and vary with operating context. Quantum error mitigation (QEM) addresses this challenge by reducing noise-induced bias at the \emph{ensemble} level -- through repeated executions, classical post-processing, and lightweight execution-time adjustments, without requiring full fault-tolerant overhead~\cite{Czarnik2021errormitigation, PhysRevA.98.062339, PhysRevA.103.042605, cai2023quantum}.

Most prior QEM work improves a specific mitigation primitive or its parameterization, such as zero-noise extrapolation, probabilistic error cancellation, and readout mitigation~\cite{cai2023quantum,temme2017error,endo2018practical}, or develops learned surrogate models that emulate a single mitigation workflow~\cite{liao2023mlqem,wang2024aiqec}. 
Recent platform and learning-based advances have further improved the cost--accuracy tradeoff of mitigation and, in some cases, enabled adaptive noise estimation under drift~\cite{daguerre2025adaptive,alexeev2024aiqc}; however, these approaches generally operate within a fixed mitigation strategy and do not address the systems problem we study here: runtime orchestration over a library of intervention options under nonstationary noise. 

ML/AI efforts have accelerated progress in quantum reliability, including learned policies for feedback and control, models that map noisy signatures to decision rules, and hybrid methods supporting calibration, compilation, and execution-time tuning~\cite{rlfosel,Sweke_2021,Varsamopoulos_2018,Baireuther_2019,Shinde2024,alexeev2024artificialintelligencequantumcomputing,wang2024artificialintelligencequantumerror}. Yet many approaches are trained and deployed as effectively \emph{static} controllers: optimized for a fixed device configuration, circuit family, or assumed noise process. In practice, static mitigation pays a constant computational ``tax'' even when noise is benign, while under-reacting when noise spikes, a mismatch that is especially costly when the appropriate response ranges from lightweight suppression (\eg dynamical decoupling, Pauli twirling) to heavier intervention (\eg code-based decoding)~\cite{Viola_1999, PhysRevA.94.052325, alexeev2024artificialintelligencequantumcomputing}.

This motivates an adaptive, cost-aware view of quantum reliability: rather than committing to a single ``best'' mitigation method, a runtime system should \emph{monitor telemetry} and \emph{choose} when to intervene and how aggressively, explicitly trading fidelity improvement against intervention cost under uncertainty. We therefore formulate QEM orchestration as a closed-loop decision problem driven by streaming telemetry, where each cycle yields an updated state summary and an immediate post-action outcome. This setting naturally admits \emph{contextual bandit} decision-making: actions are discrete, rewards are observed on a short horizon, and fast online adaptation is required without the instability and data demands of long-horizon reinforcement learning.

GSC-QEMit is differentiated by operating as a backend-agnostic policy layer above concrete mitigation primitives. 
Rather than learning a single controller tied to a specific code, circuit family, or mitigation method, it combines hierarchical telemetry context discovery, uncertainty-aware short-horizon forecasting, and cost-aware contextual bandit selection to decide \emph{when} to intervene and \emph{how aggressively} to intervene as operating conditions evolve.

To summarize, this work makes the following contributions:
\begin{itemize}
    \item \descStep{Telemetry-driven orchestration layer}{We introduce GSC-QEMit, a modular runtime framework that orchestrates mitigation intensity from streaming telemetry via hierarchical context discovery, uncertainty-aware forecasting, and a cost-aware bandit policy.}
    \item \descStep{Context--forecast factorization}{
We formalize a separation between (i) context discovery via Spark--GHSOM over streaming telemetry and
(ii) short-horizon degradation forecasting with an SVGP that provides calibrated predictive
uncertainty. This factorization enables nonstationary, context-dependent decision-making without
end-to-end retraining.
}
    \item \descStep{Cost-aware contextual bandit control}{We cast mitigation selection as contextual decision-making and implement Thompson sampling with per-action Bayesian linear models (TS-CL), bootstrapped with expert demonstrations to encode fidelity--cost preferences.}
    \item \descStep{Cross-benchmark evaluation under drift}{We validate the framework on GHZ, QFT, and Grover workloads under drifting noise in an instrumented simulator, demonstrating consistent fidelity gains and selective use of heavier interventions compared to static baselines.}
\end{itemize}

The GSC-QEMit code repo and data is available upon request.

%% file: 2026/relatedworks.tex
\vspace{-7pt}
\section{Related Work}
\label{sec:related-work}

This paper sits at the intersection of (i) quantum error mitigation (QEM) for near-term devices, (ii) learning-based quantum control and decoding, and (iii) scalable telemetry modeling for nonstationary systems. Our core distinction is architectural: rather than proposing a new mitigation primitive or a single end-to-end controller, we study \emph{mitigation orchestration}---how to select \emph{when} to intervene and \emph{which} intervention intensity to apply from streaming telemetry under explicit cost constraints.

\subsection{QEM Primitives and Overhead Trade-offs}

Canonical QEM protocols reduce noise-induced bias through repeated circuit executions and classical post-processing, including zero-noise extrapolation (ZNE) and probabilistic error cancellation (PEC)~\cite{cai2023quantum,endo2018practical}. These methods highlight a fundamental bias--variance--cost trade-off: bias reduction typically requires increased sampling overhead and/or stronger assumptions about the noise channel, which can become prohibitive as fault rates grow or as workloads scale~\cite{cai2023quantum}. This motivates approaches that adapt mitigation strength to runtime conditions, particularly under drift where the ``right'' level of mitigation may change over time~\cite{10.1145/2523813}.

\subsection{Compiler-Integrated Detection and Intermediate Mitigation}

Between purely post-processing-based QEM and full fault tolerance, quantum error detection (QED) provides a pragmatic intermediate strategy: errors are detected and runs are postselected rather than corrected, often reducing overhead relative to full correction while improving reliability. Systems work has emphasized that deploying QED at scale requires careful compilation and hardware mapping, including ancilla management and integration of detection structures into programs. QuantEM, for example, automates insertion of QED subcircuits and supports a library of detection codes as a scalable intermediate mitigation approach~\cite{liu2025quantem}. Whereas such efforts focus on \emph{compile-time} integration of detection structures, our focus is \emph{runtime adaptation}: given a library of intervention options, decide online which to enable based on streaming telemetry and forecasted degradation.

\subsection{Learning-Based Feedback Control Under Drift}

A key driver of adaptive mitigation is nonstationarity: device performance can drift due to calibration changes, environmental variation, or slow parameter drift ~\cite{doi:10.1137/1.9781611972771.42}. Prior work connects error signals (e.g., detection events or syndrome statistics) to feedback and control policies, including learned strategies that stabilize logical behavior in the presence of injected or observed drift~\cite{dong2010quantum,ahn2002continuous,rlfosel}. Reinforcement learning and related adaptive controllers further demonstrate that streaming measurements can support closed-loop control on simulated or hardware-proxied systems~\cite{rlfosel,Sweke_2021}. GSC-QEMit aligns with this telemetry-informed view, but operates at a higher policy layer: rather than directly optimizing pulse-level parameters, we select among discrete mitigation/correction \emph{intensities} using learned contexts and calibrated forecasts under explicit cost constraints.

\subsection{Neural Decoders and RL for QEC}

Substantial literature treats decoding and correction as learning problems ~\cite{Sweke_2021, Varsamopoulos_2018, battistel2023real, Shinde2024}. Supervised neural decoders map syndrome patterns to corrections with low-latency inference suitable for real-time pipelines~\cite{Varsamopoulos_2018,Baireuther_2019}, while RL-based agents can learn feedback strategies from interaction with simulated devices and streaming syndrome signals~\cite{rlfosel,Sweke_2021}. These approaches often yield high performance within a fixed code/noise setting, but they typically produce a \emph{single} policy tied to a particular code family and training distribution. In contrast, our emphasis is on \emph{orchestration}: switching between intervention options (from lighter suppression to heavier intervention) as a function of context, predicted degradation, and action cost.

\subsection{Hierarchical Context Discovery and Scalable Telemetry Modeling}

GSC-QEMit draws from unsupervised structure discovery for high-volume telemetry. Growing Hierarchical Self-Organizing Maps (GHSOM) learn coarse-to-fine partitions via data-driven growth criteria, producing interpretable hierarchical regimes rather than a single flat clustering~\cite{MALONDKAR2019572}. Spark--GHSOM extends this paradigm to distributed computation, making it a natural fit for streaming telemetry pipelines~\cite{MALONDKAR2019572,spark}. While hierarchical SOM-style models have been applied broadly to regime discovery, they have rarely been integrated with online, uncertainty-aware decision layers for quantum reliability. Our work couples Spark--GHSOM contexts to (i) a Sparse Variational Gaussian Process (SVGP) forecaster and (ii) a contextual bandit policy, enabling interpretable state descriptors, calibrated near-horizon predictions, and cost-aware adaptation under drift.

\subsection{Summary}

Across these threads, prior work has produced strong mitigation primitives (ZNE, PEC, QED), learned decoders, and feedback strategies for nonstationary noise ~\cite{PhysRevLett.119.180509, endo2018practical}. GSC-QEMit targets a complementary systems problem: given streaming telemetry, learn hierarchical operating contexts, forecast near-horizon degradation with uncertainty, and select among intervention options using a cost-aware contextual bandit.

%% file: 2026/methodology.tex
\section{Methodology}
\label{sec:methodology}
We formulate GSC-QEMit as an adaptive quantum error mitigation problem: given a stream of cycle-level telemetry, the controller must decide when to intervene and how aggressively to intervene under uncertainty and explicit intervention cost.

\begin{table}[H]
\caption{Emulator telemetry fields and provenance (per cycle).}
\centering
\scriptsize
\setlength{\tabcolsep}{3pt}
\renewcommand{\arraystretch}{1.15}
\begin{tabular}{|m{0.22\linewidth}|m{0.16\linewidth}|m{0.22\linewidth}|m{0.29\linewidth}|}
\hline
\multicolumn{4}{|c|}{\textbf{Emulator telemetry (per cycle)}} \\
\hline
\textbf{Field} & \textbf{Source} & \textbf{Derived from} & \textbf{Description} \\
\hline
\texttt{cycle} & Internal & Config & Control-cycle index $t \in \{0,\dots,T_{\mathrm{run}}-1\}$. \\
\hline
\texttt{code\_dist} & Internal & Config & Repetition distance $d\in\{3,5\}$. \\
\hline
\texttt{p\_phys} & Simulator & Noise Model & Physical noise parameter (drift knob); used to form $p_{\mathrm{eff}}$ in Table~\ref{tab:telemetry13}. \\
\hline
\texttt{eps\_logical} & Decoder & Counts & Logical error $\epsilon_L$ (TVD to ideal). \\
\hline
\texttt{fidelity} & Derived & $1-\epsilon_L$ & Logical fidelity $F_L$. \\
\hline
\texttt{features} & Derived & Current + history & Feature vector $\mathbf{x}_t \in \mathbb{R}^{13}$ (Table~\ref{tab:telemetry13}). \\
\hline
\end{tabular}
\label{tab:telemetry-provenance}
\end{table}

Rather than training a single monolithic decoder or end-to-end deep RL agent, we factor the system into three lightweight components that are trained independently and composed into a closed-loop runtime policy:
(i) \textbf{Context Discovery:} Unsupervised clustering of 13-dimensional telemetry features using Spark--GHSOM to partition execution into hierarchical operating regimes;
(ii) \textbf{Uncertainty Forecasting:} Short-horizon prediction of logical fidelity using a Sparse Variational Gaussian Process (SVGP), which provides both a predictive mean and calibrated uncertainty to guide risk-sensitive decisions; ~\cite{10.7551/mitpress/3206.001.0001}, ~\cite{hensman2015scalable} and
(iii) \textbf{Cost-Aware Decision Making:} An online linear contextual bandit policy that selects among intervention labels using Thompson Sampling with per-action Bayesian linear reward models (TS-CL). Crucially, we bootstrap the bandit via imitation learning to embed expert heuristics (\eg ``avoid heavy correction when noise is low'') before online deployment ~\cite{pmlr-v28-agrawal13,JMLR:v17:14-087}.

\paragraph{Mitigation-controller terminology.}
We define three controller behaviors used in our evaluation:
(1) Unmitigated (Identity): Applies no error mitigation, incurring zero cost but suffering full noise impact.
(2) Fixed/Static Mitigation: Applies a constant strategy (\eg Surface Code) every cycle, maximizing fidelity but maximizing cost.
(3) Adaptive Mitigation (GSC-QEMit): Dynamically selects from an intervention library $\mathcal{A} = \{\text{Identity, Pauli Suppression, Surface Code}\}$ based on the forecasted regime. In our emulator, these actions are instantiated as logical-level interventions that modify the effective noise experienced by the circuit.

\begin{table}[t]
\centering
\caption{Telemetry feature vector $\mathbf{x}_t \in \mathbb{R}^{13}$ used for context discovery and policy conditioning.
Here $t \in \{0,\dots,T_{\mathrm{run}}-1\}$ indexes the control cycle within a run of length $T_{\mathrm{run}}$.
All features are normalized to comparable scales where indicated.}
\begin{tabular}{ll}
\hline
\textbf{Idx} & \textbf{Definition} \\
\hline
1  & $t/(T_{\mathrm{run}}-1)$ \\
2  & $n/10$ \\
3  & $\mathrm{depth}/100$ \\
4  & $p_{\mathrm{eff}}$ \\
5  & $\epsilon_L$ (TVD to ideal) \\
6  & $F_L = 1-\epsilon_L$ \\
7  & $H(\hat{p})/10$ \\
8  & $N_T/50$ (T-gate count) \\
9  & $N_{\mathrm{2q}}/100$ (two-qubit gate count) \\
10 & $\log_{10}(\max(\epsilon_L,10^{-9}))$ \\
11 & $\log_{10}(\max(p_{\mathrm{eff}},10^{-9}))$ \\
12 & recent-window mean of $\epsilon_L$ \\
13 & recent-window variance of $\epsilon_L$ \\
\hline
\end{tabular}
\label{tab:telemetry13}
\end{table}

\subsection{Telemetry in Our Emulator Experiments}
\label{subsec:telemetry-emulator}
Our experiments use an instrumented logical-memory emulator that emits cycle-level telemetry. Each cycle executes a benchmark circuit (\eg GHZ, QFT) for $N_{\mathrm{shots}}$ shots. From these counts, we compute logical-memory diagnostics (detection rate, logical error rate) and assemble a fixed 13-dimensional feature vector used by all components.

\paragraph{Telemetry feature vector (13-D).}
At each control cycle $t \in \{0,\dots,T_{\mathrm{run}}-1\}$, the backend emits a
13-dimensional telemetry vector $\mathbf{x}_t$ (Table~\ref{tab:telemetry13}) used
for both hierarchical context discovery and policy conditioning. Here
$T_{\mathrm{run}}$ denotes the total number of control cycles in a run (i.e., the
experiment horizon), and time is normalized as $t/(T_{\mathrm{run}}-1)$.
The feature set combines execution progress, circuit size and depth,
gate-count statistics, and logical-performance indicators, with all quantities
scaled to comparable ranges to stabilize unsupervised clustering and linear
policy learning.

Importantly, in our simulator-based benchmark we expose
$p_{\mathrm{eff}}$ as a controlled drift knob to enable paired counterfactual
evaluation. On physical hardware, simulator-specific quantities such as
$p_{\mathrm{eff}}$ can be replaced by \emph{observable proxies} (e.g., calibration-reported
error rates, detection or syndrome event rates, readout calibration drift, or
short-window statistics of recent logical errors), while preserving the same
telemetry interface to the context and bandit layers.

\subsection{Applicability to Real Hardware}
\label{subsec:telemetry-hardware}

On real hardware, simulator-internal fields such as \texttt{p\_phys} are not directly observable. In our emulator, \texttt{p\_phys} is a controlled drift parameter used to generate noise, and we define $p_{\mathrm{eff}}$ as the scalar \emph{noise/health proxy} consumed by the learning stack (Table~\ref{tab:telemetry13}). In a physical deployment, $p_{\mathrm{eff}}$ is obtained from \emph{observable proxies} computed from periodic calibration (e.g., gate error rates, readout error, and coherence metrics $T_1$/$T_2$) and/or online indicators (e.g., syndrome/detection-event rates, leakage rates, or short-window statistics of recent logical errors). Note that $T_1$ and $T_2$ here denote standard coherence times and are unrelated to the run horizon $T_{\mathrm{run}}$ used for indexing control cycles.

\begin{algorithm}[t]
\DontPrintSemicolon
\SetAlgoLined
\footnotesize
\caption{Offline training: context $\mathcal{C}$ (GHSOM), forecaster $\mathcal{F}$ (SVGP), and bandit priors $\pi_0$}
\label{alg:offline}

\KwIn{Stored traces indexed by cycle $t \in \{0,\dots,T_{\mathrm{run}}-1\}$: features $\mathbf{f}_t \in \mathbb{R}^{13}$, fidelity $F_t$, proxy $p_{\mathrm{eff},t}$ (or simulator \texttt{p\_phys}), optional expert actions $\{a_t\}$; forecast horizon $\Delta$ (we use $\Delta=1$ in experiments).}
\KwOut{Context mapper $\mathcal{C}$; forecaster $\mathcal{F}$; bootstrapped bandit priors $\pi_0$.}

\BlankLine
\textbf{1) Context discovery (GHSOM).}\;
Compute normalization stats $(\boldsymbol{\mu}_f,\boldsymbol{\sigma}_f)$ over all $\mathbf{f}_t$\;
$\tilde{\mathbf{f}}_t \leftarrow (\mathbf{f}_t-\boldsymbol{\mu}_f)\oslash\boldsymbol{\sigma}_f$\;
Train GHSOM on $\{\tilde{\mathbf{f}}_t\}$ to obtain leaf-context labels $c_t$ and mapper $\mathcal{C}$\;

\BlankLine
\textbf{2) Uncertainty forecasting (SVGP).}\;
Form state $\mathbf{x}_t \leftarrow (c_t,\; p_{\mathrm{eff},t},\; t,\; F_t)$\;
Train SVGP on $(\mathbf{x}_t,\,F_{t+\Delta})$ to learn $p(F_{t+\Delta}\mid \mathbf{x}_t)$ and forecaster $\mathcal{F}$\;
At inference, summarize $p(F_{t+\Delta}\mid \mathbf{x}_t)$ by moments $(\hat{\mu}_{t+\Delta},\hat{\sigma}_{t+\Delta})$\;

\BlankLine
\textbf{3) Policy bootstrapping (imitation).}\;
Define cost-aware reward $r_t = F_{t+\Delta}-\lambda\,\mathrm{Cost}(a_t)$\;
Initialize linear-bandit priors $(\mathbf{A}_a,\mathbf{b}_a)$ for each action $a$\;
Construct bandit features $\mathbf{z}_t \leftarrow (\mathbf{x}_t,\hat{\mu}_{t+\Delta},\hat{\sigma}_{t+\Delta})$\;
Update priors from expert traces: $\mathbf{A}_a \leftarrow \mathbf{A}_a + \mathbf{z}_t\mathbf{z}_t^\top,\ 
\mathbf{b}_a \leftarrow \mathbf{b}_a + r_t\mathbf{z}_t$\;
Set $\pi_0 \equiv \{(\mathbf{A}_a,\mathbf{b}_a)\}_{a\in\mathcal{A}}$\;

\BlankLine
\textbf{Output:} $\mathcal{C},\,\mathcal{F},\,\pi_0$.\;
\end{algorithm}

\begin{algorithm}[t]
\DontPrintSemicolon
\SetAlgoLined
\footnotesize
\caption{Online GSC-QEMit control loop (streaming deployment)}
\label{alg:online}

\KwIn{Instrumented execution stream (backend telemetry); trained $\mathcal{C},\mathcal{F}$; bandit priors $\pi_0$; action set $\mathcal{A}$.}
\KwOut{Actions $\{a_t\}$ and logs.}

\BlankLine
Initialize bandit posterior from $\pi_0$\;

\For{$t \leftarrow 0$ \KwTo $T_{\mathrm{run}}-1$}{
\textbf{Observe:} receive telemetry $\mathbf{f}_t$ and proxy $p_{\mathrm{eff},t}$\;
\textbf{Contextualize:} $\tilde{\mathbf{f}}_t \leftarrow (\mathbf{f}_t-\boldsymbol{\mu}_f)\oslash\boldsymbol{\sigma}_f$;\;
$c_t \leftarrow \mathcal{C}(\tilde{\mathbf{f}}_t)$\;

\textbf{Forecast:} $\mathbf{x}_t \leftarrow (c_t,\; p_{\mathrm{eff},t},\; t,\; F_t)$;\;
$(\hat{\mu}_{t+\Delta},\hat{\sigma}_{t+\Delta}) \leftarrow \mathcal{F}(\mathbf{x}_t)$\;

\textbf{Decide (TS):} build $\mathbf{z}_t \leftarrow (\mathbf{x}_t,\hat{\mu}_{t+\Delta},\hat{\sigma}_{t+\Delta})$\;
Sample $\tilde{r}_a \sim \mathcal{N}(\theta_a^\top \mathbf{z}_t,\sigma^2)$ for $a\in\mathcal{A}$\;
$a_t \leftarrow \arg\max_{a\in\mathcal{A}} \left(\tilde{r}_a - \lambda\,\mathrm{Cost}(a)\right)$\;

\textbf{Act \& update:} execute $a_t$; observe $r_{t+1}$;\;
Update bandit posterior for $a_t$ using $(\mathbf{z}_t,r_{t+1})$\;
}
\end{algorithm}

Our GSC-QEMit architecture is agnostic to the provenance of the scalar health proxy: as long as a calibrated $p_{\mathrm{eff},t}$ (or equivalent) is available, the context and bandit layers can learn correlations between operating regimes and cost-effective mitigation choices.

A hardware-facing deployment runs GSC-QEMit in a batched loop: execute a monitoring batch, compute telemetry and
$p_{\mathrm{eff}}$, update context/forecast, then select an intervention for the next batch. In the emulator we instantiate
interventions as effective-noise scaling to isolate decision logic; on hardware these actions map to concrete controls
(\eg dynamical decoupling, shot allocation, measurement mitigation, or code/decoder selection where applicable).

%% file: 2026/implementation.tex
\section{Implementation}
\label{sec:implementation}

\subsection{System Architecture}
\label{sec:architecture}

\FloatBarrier
\begin{figure}[!t]
\centering
\includegraphics[width=\linewidth]{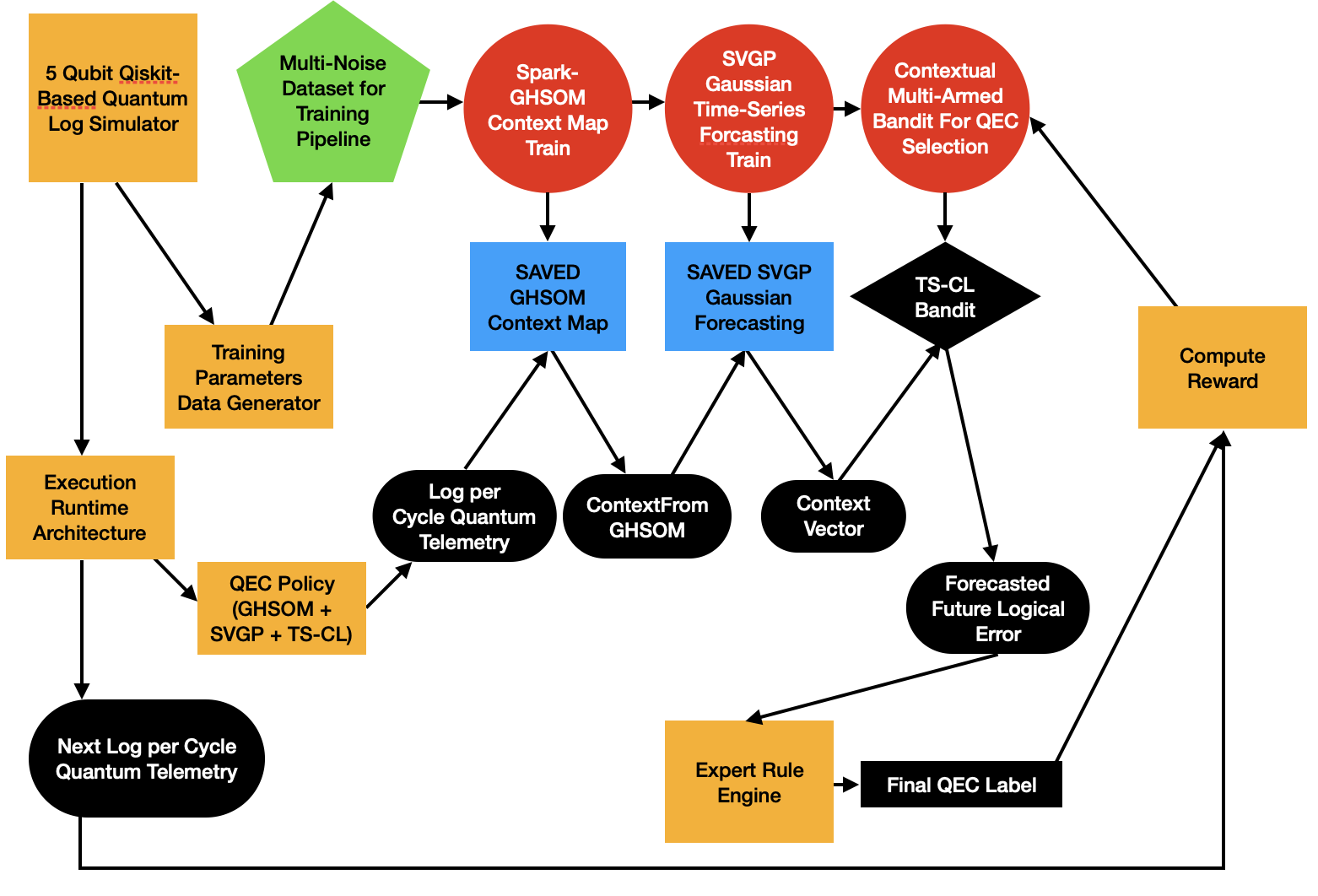}
\caption{GSC-QEMit implementation: An instrumented backend streams cycle-level telemetry to a classical context--forecast--bandit stack for adaptive mitigation selection.}
\label{fig:architecture}
\end{figure}

Figure~\ref{fig:architecture} summarizes our end-to-end implementation. An \emph{instrumented backend} provides streaming, per-cycle telemetry to a classical control stack consisting of a Spark--GHSOM context mapper, a Sparse Variational Gaussian Process (SVGP) forecaster, and a cost-aware Contextual Multi-Armed Bandit (CMAB) initialized via imitation learning. The controller is defined by a fixed telemetry interface and an abstract intervention library; the backend used here is a Qiskit Aer--based simulator for controlled evaluation, but it can be replaced by any simulator or hardware system that exposes compatible telemetry and supports a small set of intervention hooks.

\subsection{Backend Interface and Telemetry Pipeline}
\label{subsec:plant-telemetry}

We implement an instrumented testbed with a minimal control interface: a \emph{reset} operation that reinitializes time and the drift process, and a \emph{step} operation that advances the testbed by one cycle under a specified \emph{intervention level} and returns a telemetry record. This interface isolates GSC-QEMit from backend-specific details and allows the same context--forecast--policy stack to operate on any compatible telemetry stream.

\subsubsection{Intervention Library (Abstract Interface)}
GSC-QEMit is designed to sit \emph{above} any specific QEM/QED/QEC primitive. Accordingly, the controller interacts with the backend through an abstract intervention interface consisting of three mitigation-intensity levels:
\begin{itemize}
  \item \textbf{NONE}: baseline execution with no additional mitigation; zero control cost.
  \item \textbf{MODERATE}: a mid-cost intervention intended to provide partial robustness at modest overhead.
  \item \textbf{SEVERE}: a high-intensity intervention intended to maximize reliability at the highest overhead.
\end{itemize}
This abstraction is the key systems principle: different backends may map these levels to different concrete procedures (e.g., dynamical decoupling schedules, readout mitigation, noise amplification for ZNE, detection/postselection, or code-based decoding), while the context--forecast--policy stack remains unchanged.

\paragraph{Benchmark instantiation.}
In the simulator-backed benchmark used here, we instantiate these abstract levels in two complementary ways. First, \textbf{MODERATE} is realized as a structural redundancy-and-decode intervention by replicating circuit sampling and applying majority-vote decoding of measured bits to form a decoded logical distribution. Second, \textbf{SEVERE} is realized as a stronger intervention intensity via action-conditioned scaling of the effective noise parameters (and readout-flip reduction) in the Aer noise model. These choices provide a controlled test of adaptive switching under drift while preserving a general, backend-agnostic action interface.

\subsubsection{Instrumented benchmark backend with drifting noise}
The backend executes a suite of measured benchmark circuits under a time-varying noise process
simulated via Qiskit Aer. At each control cycle
$t \in \{0,\dots,T_{\mathrm{run}}-1\}$, we generate a baseline physical error rate via a bounded
sinusoidal drift around a nominal error rate $p_{\mathrm{phys}0}$ (the baseline physical error rate at the start of a run):
\begin{align}
p_{\mathrm{base}}(t)
&=
\mathrm{clip}\Big(
p_{\mathrm{phys}0}
\big[1 + \alpha \sin(2\pi t / T_{\mathrm{run}}) + \xi_t \big],
\nonumber\\
&\hspace{3em}
10^{-4},\ 0.05
\Big),
\end{align}
where $\alpha$ controls the drift amplitude, $\xi_t \sim \mathcal{N}(0,\sigma^2)$ models
stochastic fluctuations, and $\mathrm{clip}(\cdot)$ enforces physically reasonable bounds.

An intervention-dependent scale factor $s(a)$ is then applied to obtain the effective physical
error probability:
\begin{equation}
p_{\mathrm{eff}}(t,a)
=
\mathrm{clip}\!\left(
p_{\mathrm{base}}(t)\cdot s(a),
\ 10^{-5},\ 0.5
\right),
\end{equation}
where $a\in\mathcal{A}=\{\texttt{NONE},\texttt{MODERATE},\texttt{SEVERE}\}$ denotes the selected
intervention level. Stronger intervention levels correspond to smaller $s(a)$ (greater error
suppression) but incur higher control cost $C(a)$ in the decision layer.

Importantly, in this simulator-based benchmark we expose $p_{\mathrm{eff}}$ as a controlled drift
knob for repeatable evaluation. On physical hardware, simulator-specific quantities such as
$p_{\mathrm{eff}}$ can be replaced by \emph{observable proxies} (e.g., calibration-reported error
rates, detection/syndrome event rates, readout drift indicators, or short-window statistics of
recent errors) without changing the controller interface.

\subsubsection{Telemetry logging schema}
Following the QDataSet design philosophy~\cite{perrier2022qdataset}, we store each control cycle as a JSON object containing inputs, metadata, and derived diagnostics. Key fields include:
\begin{itemize}
  \item \texttt{"cycle"}: logical round index $t$.
  \item \texttt{"p\_eff"}: effective physical error proxy $p_{\mathrm{eff}}(t,a)$ (simulator-controlled in this benchmark).
  \item \texttt{"eps\_logical"}: logical error $\epsilon_L$ computed as the TVD between the observed and ideal output distributions.
  \item \texttt{"fidelity\_logical"}: logical fidelity $F_L = 1 - \epsilon_L$.
  \item \texttt{"features"}: a 13-D feature vector $\mathbf{f}_t$ used for context mapping (Table~\ref{tab:telemetry13}).
  \item \texttt{"action\_level"}: selected intervention level in \{\texttt{NONE}, \texttt{MODERATE}, \texttt{SEVERE}\}.
\end{itemize}

\subsection{Spark--GHSOM Context Mapping}
\label{subsec:spark-ghsom}

We train a Growing Hierarchical Self-Organizing Map (GHSOM) on the 13-D features using Apache Spark. The objective is to discretize streaming telemetry into hierarchical operating regimes (contexts) that remain informative under drift. Training yields a context index $\mathcal{C}: \mathbb{R}^{13} \to \mathbb{Z}$ that maps each normalized feature vector $\tilde{\mathbf{f}}_t$ to a unique discrete context identifier $c_t$ used at runtime.

\subsection{Sparse Variational GP (SVGP) Forecaster}
\label{subsec:svgp}

To predict near-horizon degradation with calibrated uncertainty, we employ a Sparse Variational
Gaussian Process (SVGP). Unlike standard GPs, which scale cubically in the number of training
points ($O(N^3)$), SVGP uses a set of inducing points to form a variational approximation to the
posterior, enabling efficient training and fast runtime querying.

At each cycle $t$, the forecaster consumes a compact, interpretable state vector
\[
\mathbf{x}_t = \big[c_t,\ p_{\mathrm{eff},t},\ d,\ t/T_{\mathrm{run}},\ F_{L,t}\big],
\]
where $c_t$ is the discrete context label output by the GHSOM, $p_{\mathrm{eff},t}$ is the current
scalar noise/health proxy, $d$ is the code distance of the executed circuit, $t/T_{\mathrm{run}}$ is
normalized execution time, and $F_{L,t}$ is the observed logical fidelity. Conditioned on
$\mathbf{x}_t$, the SVGP returns a Gaussian predictive distribution over the near-horizon fidelity:
\begin{equation}
F_{L,t+\Delta} \sim \mathcal{N}\!\big(\hat{\mu}_{t+\Delta},\,\hat{\sigma}^2_{t+\Delta}\big),
\end{equation}
where $(\hat{\mu}_{t+\Delta}, \hat{\sigma}_{t+\Delta})$ are the posterior mean and standard deviation.

Rather than sampling from this distribution, GSC-QEMit treats these moments as summary statistics
of predicted system health and appends them deterministically to the bandit context vector,
\[
\mathbf{z}_t = \big[\mathbf{x}_t,\ \hat{\mu}_{t+\Delta},\ \hat{\sigma}_{t+\Delta}\big].
\]
The contextual bandit then applies Thompson sampling over its own reward model using $\mathbf{z}_t$,
so that action selection depends on both expected future fidelity and forecast uncertainty, enabling
a performance--risk trade-off when choosing mitigation intensity.

\subsection{Cost-Aware Contextual Bandit (TS-CL)}
\label{subsec:bandit}

Adaptive mitigation is implemented as a Contextual Multi-Armed Bandit (CMAB) using Thompson Sampling with Linear Payoffs (TS-CL). At cycle $t$, the bandit context vector is formed by appending the SVGP forecast moments to the current compact state:
\begin{align}
  \mathbf{z}_t
  =
  \big[\mathbf{x}_t,\ \hat{\mu}_{t+\Delta},\ \hat{\sigma}_{t+\Delta}\big]^\top
\end{align}
For each intervention level $a\in\mathcal{A}$, we maintain a Bayesian linear reward model characterized by a multivariate normal posterior over weight vectors $\boldsymbol{\theta}_a$:
\begin{equation}
  \boldsymbol{\theta}_a \sim \mathcal{N}(\hat{\boldsymbol{\theta}}_a, \mathbf{\Sigma}_a).
\end{equation}
We sample $\tilde{\boldsymbol{\theta}}_a$ from the posterior to obtain a stochastic value estimate for each action.

\subsubsection{Cost-Aware Reward Function}
Crucially, the objective is not just to maximize fidelity, but to maximize \emph{value}: reliability improvement per unit cost. We define the realized reward as improvement in logical error minus a penalty for intervention overhead:
\begin{equation}
  r_t(a) = \big(\epsilon_{L,t}-\epsilon_{L,t+1}\big) - \lambda \cdot C(a),
  \label{eq:reward-function}
\end{equation}
where $C(a)\in[0,1]$ is the normalized computational/control cost of intervention level $a\in\{\texttt{NONE},\texttt{MODERATE},\texttt{SEVERE}\}$, and $\lambda$ is a tunable Lagrange multiplier controlling the frugality of the agent.

\subsubsection{Imitation Learning (Behavioral Cloning)}
To overcome the cold-start problem where an uninformative prior may fail to discover effective switching behavior within a finite horizon, we bootstrap the bandit using offline imitation learning. We generate synthetic traces of an expert heuristic (e.g., selecting \texttt{NONE} under low inferred noise and \texttt{SEVERE} under high inferred noise) and perform Bayesian updates on the bandit's priors $(\hat{\boldsymbol{\theta}}_a, \mathbf{\Sigma}_a)$ before online deployment. This seeds the agent with domain knowledge while allowing it to fine-tune its behavior online.

\subsection{Runtime Orchestrator}
\label{subsec:runtime}

The runtime orchestrator executes closed-loop control cycle:
\begin{enumerate}
  \item \textbf{Observe:} receive telemetry $\mathbf{f}_t$ from the backend.
  \item \textbf{Contextualize:} map features to GHSOM context $c_t$ and query SVGP for forecast $(\hat{\mu}_{t+1}, \hat{\sigma}_{t+1})$.
  \item \textbf{Decide:} build $\mathbf{z}_t$, sample action values via Thompson Sampling, subtract costs, and select the optimal intervention level $a_t$.
  \item \textbf{Act:} execute $a_t$ on the backend, observe the new state and error $\epsilon_{L,t+1}$, and compute the realized reward $r_t$.
  \item \textbf{Update:} update the posterior distributions for the selected intervention level $a_t$.
\end{enumerate}
This modular implementation supports direct comparisons among \textbf{unmitigated}, \textbf{static}, and \textbf{adaptive} strategies, isolating the specific value of dynamic switching under drift.

%% file: 2026/results.tex
\section{Results}
\label{sec:results}

We evaluate GSC-QEMit on a suite of eight benchmark circuit workloads spanning three families: \textbf{Clifford} (e.g., GHZ state, Bell chain), \textbf{Non-Clifford} (e.g., CCX-heavy, $T$-gate sweep), and \textbf{Structured Algorithms} (e.g., QFT, Grover, Bernstein--Vazirani). All experiments use an instrumented Qiskit Aer backend with a controlled nonstationary drift process; the effective physical error rate varies sinusoidally over time (implemented through the simulator noise model) to emulate changing operating conditions.

\begin{figure}[t]
    \centering
    \includegraphics[width=1\linewidth]{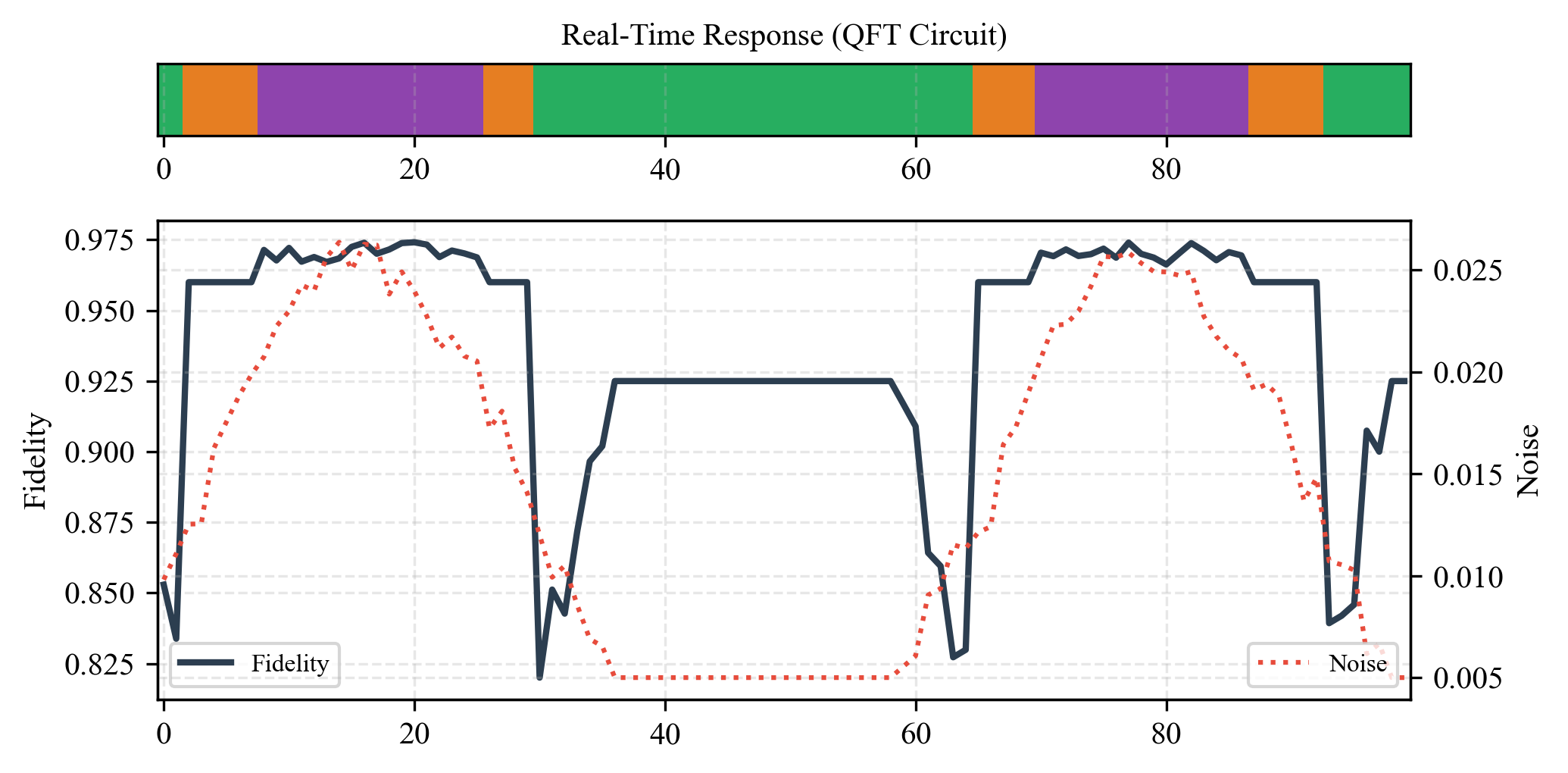}
    \caption{\textbf{Real-time adaptive response (case study: QFT).} A representative execution trace under drifting noise.
    \textbf{Top strip (barcode):} the selected intervention level over time (\texttt{NONE} $\rightarrow$ \texttt{MODERATE} $\rightarrow$ \texttt{SEVERE}).
    \textbf{Barcode shading encodes action: white/green = NONE, gray/orange = MODERATE, black/purple = SEVERE.}
    \textbf{Bottom plot:} logical fidelity $F_L$ (solid) alongside the drifting effective noise indicator (dotted).
    \textbf{Key insight:} during elevated-noise intervals, the policy increases intervention intensity (switching to \texttt{SEVERE}); during quiescent periods it returns to \texttt{NONE}/\texttt{MODERATE}, conserving cost while maintaining high fidelity.}
    \label{fig:barcode}
\end{figure}

\begin{figure*}[t]
    \centering
    \includegraphics[width=1\linewidth]{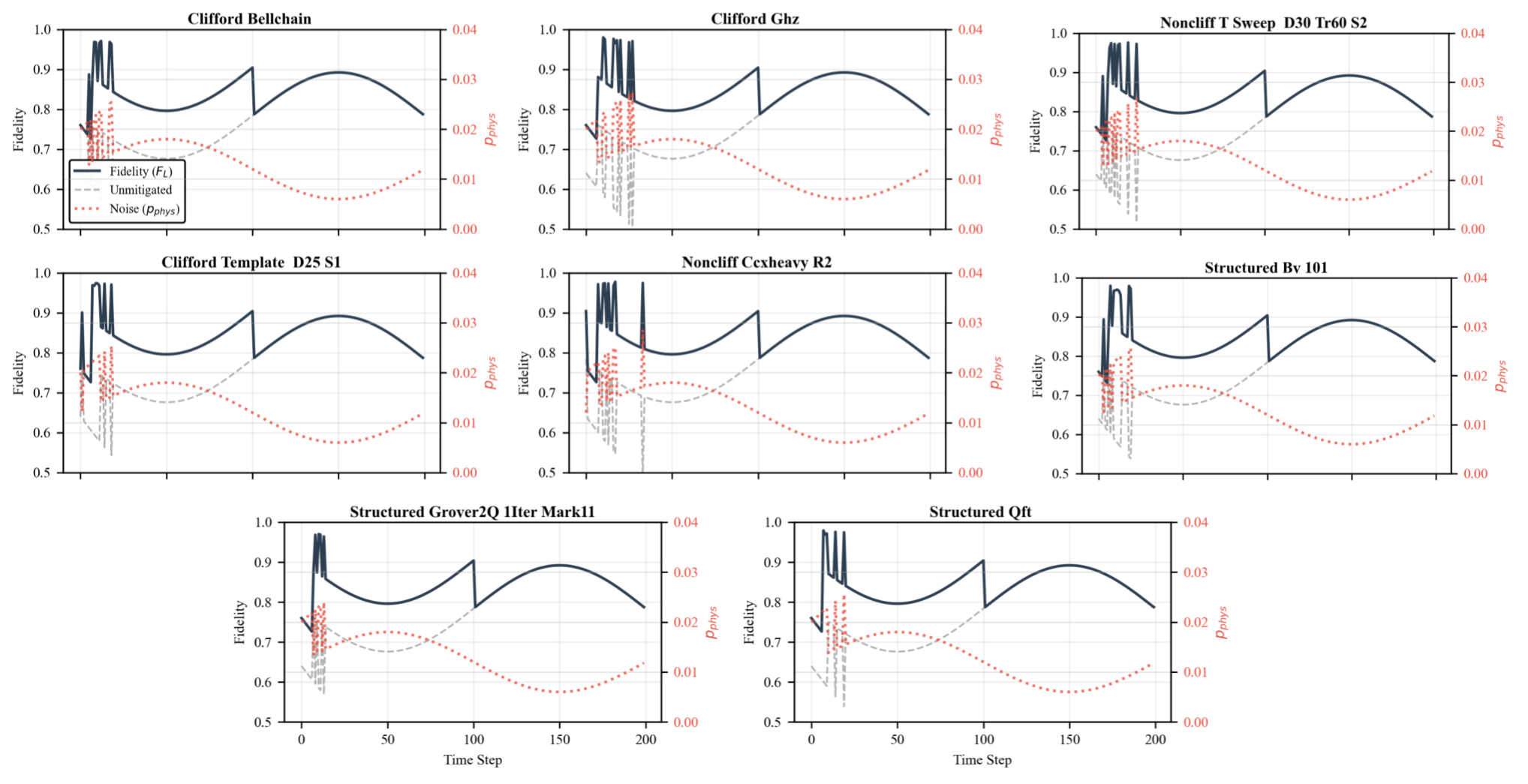}
    \caption{\textbf{Cross-benchmark stabilization under drift.}
    Time-series traces for the evaluation suite under a controlled, time-varying noise schedule with a deliberate mid-run noise peak.
    \textbf{Dotted:} imposed effective noise indicator $p_{\mathrm{eff}}(t)$.
    \textbf{Dashed:} unmitigated baseline logical fidelity.
    \textbf{Solid:} adaptive fidelity under GSC-QEMit.
    The reduced depth of the mid-run fidelity trough under the adaptive policy reflects active
    mitigation escalation during high-noise intervals, followed by de-escalation as noise relaxes.}
    \label{fig:traces}
\end{figure*}

\noindent We compare three strategies under identical drift realizations:
\begin{enumerate}
    \item \textbf{Unmitigated (Baseline):} No intervention is applied (equivalent to intervention level \texttt{NONE} at every cycle).
    \item \textbf{Static Severe:} A fixed high-intensity intervention is applied at every cycle (equivalent to \texttt{SEVERE} at all times), incurring constant overhead.
    \item \textbf{Adaptive (GSC-QEMit):} The agent selects among \texttt{NONE}, \texttt{MODERATE}, and \texttt{SEVERE} each cycle based on telemetry, forecasted near-horizon degradation, and an explicit cost penalty.
\end{enumerate}

\subsection{Real-Time Stabilization Under Drift}
\label{subsec:stabilization}

Figure~\ref{fig:traces} illustrates representative telemetry traces for the full benchmark suite
under a controlled, time-varying noise process. In all the experiments, circuits are executed repeatedly over a fixed horizon $T_{\mathrm{run}}$ while the effective physical error rate $p_{\mathrm{eff}}(t)$ is deliberately modulated according to the drifting-noise model described in
Section~\ref{subsec:plant-telemetry}. No additional faults or adversarial perturbations are injected
beyond this explicit drift process.

The drift schedule is designed to probe two complementary regimes. In the first half of each run,
the noise level is rapidly increased to a peak, producing a pronounced high-noise interval near the
center of the trace. This stress-test phase evaluates whether the controller can detect elevated risk
and escalate mitigation quickly. In the second half of the run, the noise level varies more slowly
and gradually relaxes, allowing us to observe whether the policy correspondingly de-escalates
intervention when aggressive mitigation is no longer beneficial.

Across all six benchmarks, the unmitigated baseline (dashed curves) exhibits a clear fidelity
collapse during the high-noise interval, producing the characteristic mid-run trough visible in each
panel. In contrast, the adaptive controller (solid curves) responds to the rising predicted risk by
escalating intervention intensity, substantially reducing the depth of these troughs. As the noise
later decreases and stabilizes, the controller relaxes mitigation, causing the adaptive fidelity to
converge toward the unmitigated trace. This convergence reflects intentional behavior: when noise is
low and stable, the optimal policy favors minimal intervention to avoid unnecessary overhead.

Overall, the reduced sensitivity of the adaptive fidelity traces to the imposed noise peak
demonstrates a genuine stabilization effect. Rather than merely tracking drift, GSC-QEMit actively
counteracts time-varying noise by modulating mitigation strength in response to forecasted system
health.

\subsection{Quantitative Performance Across Workloads}
\label{subsec:generalization}

To summarize performance across workloads, Table~\ref{tab:benchmark_gain} reports mean fidelity under unmitigated execution and under the adaptive policy, along with relative gain
$\Delta(\%) = 100 \times \left(\frac{F_L^{\mathrm{adaptive}}}{F_L^{\mathrm{unmitigated}}}-1\right)$.

\begin{table}[htbp]
\caption{Performance by benchmark. We report mean unmitigated fidelity, mean adaptive fidelity under GSC-QEMit, and relative gain $\Delta(\%)$.}
\vspace{-10pt}
\begin{center}
\begin{tabular}{|c|c|c|c|c|}
\hline
\multicolumn{5}{|c|}{\textbf{Benchmark Performance Summary}} \\
\hline
\textbf{\textit{Class}} & \textbf{\textit{Benchmark}} & \textbf{\textit{Unmitigated}} & \textbf{\textit{Adaptive}} & \textbf{\textit{Gain (\%)}} \\
\hline
Clifford      & Bell Chain        & 0.774 & 0.843 & \textbf{+8.9\%} \\
Clifford      & GHZ State         & 0.771 & 0.844 & \textbf{+9.4\%} \\
Clifford      & Template          & 0.774 & 0.844 & \textbf{+9.1\%} \\
\hline
Non-Clifford  & CCX-Heavy         & 0.773 & 0.844 & \textbf{+9.2\%} \\
Non-Clifford  & T-Sweep           & 0.773 & 0.844 & \textbf{+9.1\%} \\
\hline
Structured    & BV (Oracle)       & 0.773 & 0.845 & \textbf{+9.3\%} \\
Structured    & Grover (Search)   & 0.775 & 0.841 & \textbf{+8.5\%} \\
Structured    & QFT (Algo)        & 0.774 & 0.842 & \textbf{+8.7\%} \\
\hline
\end{tabular}
\label{tab:benchmark_gain}
\end{center}
\end{table}

Across all benchmarks, GSC-QEMit yields consistent improvements in mean logical fidelity (\textbf{+8.5\% to +9.4\%}) relative to unmitigated execution. Gains are comparable across the three workload families, suggesting that the telemetry-driven interface captures degradation patterns that are not tied to a single circuit type. 

Finally, the adaptive policy achieves these fidelity gains while reducing unnecessary high-intensity intervention. In our runs, the controller selects \texttt{NONE} during low-risk intervals for approximately 40\% in the reported setting, thereby reducing aggregate intervention cost by $\sim$35\% compared to high-intensity intervention at every cycle.

%% file: conclusion.tex
\section{Conclusion}
\label{sec:conclusion}

We present \emph{GSC-QEMit}, a telemetry-driven framework for adaptive quantum error mitigation that formulates mitigation selection as a cost-aware policy under nonstationary drift. The approach composes three independently trained, lightweight modules: a Spark--GHSOM context mapper, an SVGP forecaster that predicts near-horizon fidelity with calibrated uncertainty, and a TS-CL contextual bandit that selects among abstract intervention levels (\texttt{NONE}, \texttt{MODERATE}, \texttt{SEVERE}).

Evaluated on an instrumented Qiskit Aer benchmark suite spanning eight circuit workloads, GSC-QEMit stabilizes logical fidelity by escalating mitigation during high-risk intervals and relaxing it under benign conditions. Across workloads, it improves mean logical fidelity by \textbf{+9.0\%}, reducing aggregate intervention cost by about 35\% relative to a static severe strategy. Results show that combining context discovery with uncertainty-aware forecasting enables robust, budgeted mitigation that transfers across workloads without circuit-specific tuning.

\section*{Acknowledgements}
This material is based upon work supported by the United States National Science Foundation (\#2225354). 